# BanglaMATH : A Bangla benchmark dataset for testing LLM mathematical reasoning at grades 6, 7, and 8

Tabia Tanzin Prama[1,2,3,5], Christopher M. Danforth[1,2,3,4], Peter Sheridan Dodds[1,2,3,5,6]

[1]Computational Story Lab, [2]Vermont Complex Systems Institute,
[3]Vermont Advanced Computing Center,
[4]Department of Mathematics and Statistics, [5]Department of Computer Science,
University of Vermont, Burlington, VT 05405, USA
[6]Santa Fe Institute, 1399 Hyde Park Rd, Santa Fe, NM 87501, USA

## Abstract

Large Language Models (LLMs) have tremendous potential to play a key role in supporting mathematical reasoning, with growing use in education and AI research. However, most existing benchmarks are limited to English, creating a significant gap for low-resource languages. For example, Bangla is spoken by nearly 250 million people who would collectively benefit from LLMs capable of native fluency. To address this, we present BanglaMATH, a dataset of 1.7k Bangla math word problems across topics such as Arithmetic, Algebra, Geometry, and Logical Reasoning, sourced from Bangla elementary school workbooks and annotated with details like grade level and number of reasoning steps. We have designed BanglaMATH to evaluate the mathematical capabilities of both commercial and open-source LLMs in Bangla, and we find that Gemini 2.5 Flash and DeepSeek V3 are the only models to achieve strong performance, with $\geq 80\%$ accuracy across three elementary school grades. Furthermore, we assess the robustness and language bias of these top-performing LLMs by augmenting the original problems with distracting information, and translating the problems into English. We show that both LLMs fail to maintain robustness and exhibit significant performance bias in Bangla. Our study underlines current limitations of LLMs in handling arithmetic and mathematical reasoning in low-resource languages, and highlights the need for further research on multilingual and equitable mathematical understanding. Dataset link: https://github.com/BanglaMATH

## 1 Introduction

Mathematical reasoning is one of the cornerstones of human intelligence and remains a critical focus in the pursuit of artificial intelligence (AI). As AI continues to evolve, empowering machines with a deep and comprehensive understanding of mathematics not only showcases technological advancement but also represents a key milestone toward developing more generalized and capable AI systems. Recently, the emergence of Large Language Models (LLMs) has significantly reshaped the AI landscape, establishing them as powerful tools for automating complex tasks. LLMs have demonstrated remarkable proficiency in mathematical problem-solving (Romera-Paredes et al., 2023; Imani et al., 2023), prompting extensive evaluations of their capabilities across a variety of domains (Liu et al., 2023; Deng et al., 2024; Wu et al., 2023). Notably, LLMs such as ChatGPT (Ouyang et al., 2022) along with (Taylor et al., 2022) have shown impressive performance in generating and interpreting natural language, while the more recent GPT-4 (OpenAI, 2023; Bubeck et al., 2023) has set new standards in both linguistic and logical tasks.

The capacity to understand and solve mathematical problems is an especially desirable trait for LLMs, with wide-ranging applications in education, science, and industry. However, evaluating their mathematical competence is inherently challenging. While numerous benchmark datasets have been developed to assess mathematical reasoning (Cobbe et al., 2021; Amini et al., 2019; Hendrycks et al., 2021), most are limited to the English language, with a few available in Chinese (Zhang et al., 2023a,b; Zhao et al., 2020; Zhou et al., 2023). Mathematical reasoning in Bangla, despite the widespread use of LLMs by Bangla-speaking communities for educational purposes, remains largely unexplored.

To address this gap, we introduce the Bangla Mathematical Benchmark (BanglaMATH) dataset which consists of 1.7k elementary school level math problems from Bangladesh, the first benchmark dataset designed specifically to evaluate mathematical reasoning in Bangla. The BanglaMATH dataset comprises 1.7k elementary-level math word problems sourced from authentic Bangla workbooks and examination materials. The dataset includes multiple-choice questions, logical puzzles, and descriptive reasoning problems, each annotated with question type metadata to facilitate fine-grained evaluation. In Figure 1, we display an example problem from the dataset, in Bangla and English, with the correct (human) answer and and an incorrect one from ChatGPT. We posit that the evaluation of LLMs should mirror that of human learners, which would allow us to convey results in a manner that is more intuitive and accessible.

We assess the performance of several widely used Large Language Models (LLMs)—including both commercial APIs and open-source models—on the BanglaMATH dataset. Each problem in BanglaMATH is annotated with grade-level information, allowing us to conduct fine-grained evaluations similar to stating "ChatGPT scored 72 out of 100 in a sixth-grade math exam."

Our results show that Gemini 2.5 Flash and DeepSeek V3 consistently achieve high performance (accuracy $\geq$ 80%) across all three elementary school grade levels. We evaluate model performance across varying levels of arithmetic and reasoning complexity, and observe that accuracy significantly decreases as problem complexity increases. We also examine the robustness of these models by injecting distracting information into the math problems. Our findings reveal that both of the top-performing models (Gemini 2.5 Flash and DeepSeek V3) are easily misled by the presence of such irrelevant information, resulting in incorrect reasoning and answers. To further investigate language bias in LLMs, we translate the BanglaMATH dataset into English and re-evaluate the top-performing models (Gemini 2.5 Flash and DeepSeek V3). Interestingly, both models show an improvement in accuracy by $\geq$ 6.5% when tested on the English-translated version, highlighting a performance disparity for low-resource languages like Bangla.

## 2 BanglaMATH Dataset

We can frame our work as being driven by the question:

> Do LLMs have the ability to perform mathematical reasoning in Bangla?

To address this, we introduce the BanglaMATH dataset, specifically designed to evaluate LLMs on a diverse set of mathematical problems written in Bangla. Our focus spans across various categories of mathematics, including Arithmetic, Algebra, Geometry, and Logical Reasoning, to provide a comprehensive assessment of LLMs' general-purpose reasoning and arithmetic capabilities.

### 2.1 Data Collection

The mathematical question dataset is compiled from a wide range of elementary school level sources, such as school exercise books, quizzes, and exam equations. The original materials are primarily in PDF or Microsoft Word formats. These were converted into plain text using automated tools where possible, with manual transcription performed by human annotators when necessary. Since our interest lies solely in text-based word problems, we excluded any questions requiring visual or graphic interpretation. All collected questions undergo a rigorous pre-processing pipeline, including de-duplication and cleaning. This is followed by multiple rounds of human validation by the authors to ensure the quality and accuracy of the dataset.

**Data annotation.** Each problem in the dataset is annotated with several attributes, including grade level, answer, number of reasoning steps, explanations, and the number of digits involved. We provide a statistical summary of the BanglaMATH benchmark dataset in Table 1.

| Grade | Size | Steps | Digits | Length |
| --- | --- | --- | --- | --- |
| Eight | 516 | 2.19 | 2.26 | 10.17 |
| Seven | 679 | 1.82 | 2.09 | 8.99 |
| Six | 508 | 1.69 | 1.41 | 7.78 |

Table 1: Statistics of the BanglaMATH dataset. "Length" denotes the average problem length in terms of the number of words. "Steps" indicates the average number of reasoning steps required to solve a problem. "Digits" refers to the average number of digits involved in each problem's solution.

**Grade.** Each math problem in the dataset is annotated with its corresponding elementary school grade. This information was collected while gathering subsets of problems specific to each grade, enabling more targeted and accurate grade-level evaluation.

**Ground Truth Answer.** For each problem, we also annotate the ground-truth answer. Most of the answers are standalone numericals—integers, decimals, fractions, or percentages. A small proportion of questions, however, have descriptive answers. Among all problems, 9.63% require descriptive answers, while 90.37% can be answered with a single-word numerical response. We do not provide the reasoning process leading to the answer, as our dataset is intended solely for test purposes.

**Number of Steps.** Additionally, for each problem, we manually annotate the number of reasoning steps required to solve it. For the majority of problems, human annotators can easily reach consensus on the number of steps. In a few cases, the number of steps may vary depending on the specific solution considered, but this ambiguity affects only a small fraction of the dataset and should not pose a significant issue. We use the number of reasoning steps as a proxy for a problem's complexity, which reflects the level of logical analysis and problem-solving strategies required for a language model to arrive at the correct solution. Generally, more reasoning steps correspond to a more intricate thought process and provide more opportunities for the model to make errors or lose track of the problem's structure.

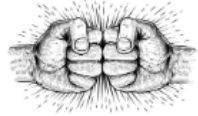

Figure 1: An example from the BanglaMATH dataset illustrating a discrepancy between the correct human-provided explanation and the incorrect response generated by ChatGPT-4o (as of May 18, 2025).

**Number of Digits.** To determine the number of digits, we identify the maximum base-10 length of any number appearing in the problem using Algorithm 1. Only digits 0–9 are counted; other symbols, such as decimal points, percent signs, and slashes, are ignored.

---
**Algorithm 1** Compute Maximum Digit Count in a Math Word Problem
---
1: **function** max_digit_count(problem_statement, answer)
2:     $N \leftarrow$ extract all numbers from *problem_statement* as strings
3:     Append str(*answer*) to $N$
4:     digit_counts $\leftarrow [\,\text{len}(x)$ for each $x \in N\,]$
5:     $D \leftarrow \max(\text{digit\_counts})$
6:     **return** $D$
7: **end function**
---

For example, in the problem statement shown in Figure 1, the maximum number is 29, so the digit count is 2. Table 2 shows the annotated samples of BanglaMATH dataset.

## 3 Experimental Setup

### 3.1 Models

We evaluate a range of widely used LLMs capable of processing Bangla text and fine-tuned for general-purpose reasoning tasks. These models, developed by various organizations, differ in architecture, size, and access methods—some are available via APIs, others through open-source model weights.

- GPT-4 (OpenAI, 2023) (released on March 14, 2023) is a multimodal LLM that demonstrates human-level performance across a range of professional and academic benchmarks. Based on the Transformer architecture, it is pre-trained to predict the next token in a sequence and is capable of analyzing, reading, and generating up to 25,000 words (32,768 tokens) per input. The model is estimated to have 1.76 trillion parameters which is accessible via ChatGPT Plus and the OpenAI API. In this experiment, we utilized the OpenAI API to access and evaluate the GPT-4 model.

- LLaMA 4 (AI, 2025) (released April 5, 2025) is the latest model from Meta AI designed to enable more personalized and natively multimodal experiences. LLaMA 4 is available in two main versions: Llama 4 Scout, a 17-billion active parameter model with 16 experts, and Llama 4 Maverick, a 17-billion active parameter model with 128 experts. Llama 4 Maverick is regarded as the best multimodal model in its class, outperforming GPT-4o and Gemini 2.0 Flash and DeepSeek v3 in reasoning and coding tasks. In this experiment, we evaluate the Llama 4 Maverick model using the Meta.AI [1] website.

Table 2: Sample problems with their English translations (not part of the dataset) and human annotations. The columns "Answer," "#Steps," and "#Digits" refer to the ground truth answer, number of reasoning steps, and the maximum number of digits in the problem, respectively

| Grade | Question (Bangla) | Question (English) | Answer (Bangla) | Answer (English) | #Steps | #Digits |
|---|---|---|---|---|---|---|
| 6 | ডাক্তার এক রোগিকে সকাল ৬টা থেকে দুপুর ১২টা পর্যন্ত আধ ঘন্টা পরপর ১টি টেবলেট খেতে বললেন। রোগীর মোট কতটি টেবলেট প্রয়োজন? | The doctor asked a patient to take 1 tablet every half hour from 6 AM to 12 PM. How many tablets does the patient need in total? | ১৩ টা | 13 | 2 | 2 |
| 7 | একটি ৫০০ মিটার লম্বা ট্রেনের গতি ৬০ কিলোমিটার হলে ,অধকিলোমীটার লম্বা একটি সেতু পাড়ি দিতে ট্রেনটির কতক্ষণ সময় লাগবে? | If the speed of a 500 m long train is 60 km/h, how much time will it take to cross a half km long bridge? | ১ মিনিট | 1 minute | 2 | 3 |
| 8 | প্রথম সাতটি মৌলিক সংখ্যার গড় নির্ণয় কর। (ক) 5.60 (খ) 8.28 (গ) 7.42 (ঘ) 6.84 | Find the average of the first seven prime numbers. (a) 5.60 (b) 8.28 (c) 7.42 (d) 6.84 | খ | B | 3 | 1 |
| 8 | যখন একটি আর্টিকেল 20% লাভে বিক্রি হয়, তখন এটি থেকে যখন এটি 20% হারে বিক্রি হয় তার চেয়ে 60 টাকায় বেশি বিক্রি হয়। আর্টিকেলটির দাম কত? (ক) 150 টাকা (খ) 200 টাকা (গ) 140 টাকা (ঘ) 120 টাকা | When an article sells at 20% profit, it sells for Rs.60 more than when it sells at 20%. How much does the article cost? (a) Rs.150 (b) Rs.200 (c) Rs.140 (d) Rs.120 | ক | A | 3 | 3 |
| 6 | কোনো সংখ্যা অক্ষের সাহায্যে লেখাকে অঙ্কপাতন বলে? | What is Notation? | কোনো সংখ্যা অক্ষের সাহায্যে লেখাকে অঙ্কপাতন বলে | system of written symbols used to represent numbers | 1 | - |

- Gemini 2.5 Flash (Gemini Team, 2025) (released June 17, 2025) is the latest model from Google, featuring a maximum input of 1,048,576 tokens and a default maximum output of 65,535 tokens and built on a sparse mixture-of-experts (MoE) (Clark et al., 2022) Transformer architecture. It incorporates advanced reasoning capabilities, allowing users to observe the model's "thinking process" as it generates responses. Gemini 2.5 Flash also introduces agentic AI, supports real-time applications, and is optimized for large context processing with up to 1 million input tokens. The model is accessible through Google AI Studio and Vertex AI. For this experiment, we eval-uated Gemini 2.5 Flash using the Google AI Studio [2] platform.

---
[1] https://ai.meta.com/

- Grok 3 (released February 17, 2025) is the latest AI model from xAI, combining transformer-based language modeling with symbolic reasoning modules in a 1.2 trillion parameter architecture (Inaba et al., 2003). Grok 3 employs 128 expert networks with dynamic routing, enabling specialized processing for different task types while maintaining 83% parameter activation efficiency (Doshi et al., 2023). Unlike traditional mixture-of-experts (MoE) models, Grok 3 introduces cross-expert attention gates, allowing knowledge sharing between specialized components without catastrophic interference. The training corpus comprises 13.4 trillion tokens which is accessible via official website and X. For this experiment, we use Grok 3 through official website[3].

- DeepSeek-V3 (DeepSeek-AI et al., 2024) (released December 26, 2024) is a Mixture-of-Experts (MoE) language model with 671 billion total parameters, of which 37 billion are activated for each token. DeepSeek-V3 utilizes Multi-head Latent Attention (MLA) and the DeepSeekMoE architecture, building on advances from DeepSeek-V2, to enable efficient inference and cost-effective training. The model is pre-trained on 14.8 trillion tokens, and optimized through supervised fine-tuning and reinforcement learning. It can be accessed directly via chat on the official website and API through the DeepSeek Platform. For this experiment, we evaluated DeepSeek-V3 using the official website.

### 3.2 Evaluation Procedure

In our experiments, we employ a zero-shot evaluation approach, refraining from using any auxiliary prompting strategies. Each math problem is presented to the LLMs in its original Bangla text format, without any additional context, examples, or instructions, as shown in Figure 2.

We deliberately choose zero-shot evaluation to reflect a realistic and practical deployment scenario. Since the LLMs considered in this study are fine-tuned for general-purpose use and are designed to function out-of-the-box, we argue that zero-shot evaluation offers a more reliable measure of their baseline capabilities.

---

[2]https://deepmind.google/
[3]https://grok.com

**Direct Prompting (DP)**

তোমাকে একটি গণিত প্রশ্ন দেওয়া হবে বাংলায়।
প্রশ্নটি বুঝে শুধুমাত্র সেই প্রশ্নের উত্তর বাংলায় দাও।
ব্যাখ্যা, ধাপ লিখবে না — কেবলমাত্র প্রশ্নের উত্তর বাংলায় দাও।

প্রশ্ন: {question}
উত্তর: {Answer}

**Translated Prompt**

You will be given a math question in Bengali.
Understand the question and provide only the answer in Bengali.
Do not write any explanation—only give the answer

Question: {question}
Answer: {Answer}

Figure 2: Example prompt provided to the LLMs. Note that Bengali is an alternative name for the language Bangla.

Finally, we calculate accuracy scores as part of our evaluation. To assess performance, we extract the numerical answer(s) generated by the model and compare with the annotated ground truth answers. We also conduct manual verification of the LLM-generated outputs and their corresponding ground truth answers to ensure the accuracy of our evaluation, and to account for possible formatting inconsistencies or ambiguities in the responses. An LLM response is marked correct if there is an exact numerical match for single-word numerical responses, or if descriptive answers are judged to be similar in meaning.

## 4 Result and Analysis

### 4.1 Grade level accuracy

The test results [4] are presented in Figure 3, illustrating the accuracy of each model. A notable downward trend in accuracy can be observed, indicating that the performance of all models declines when grade level is increased. Although this outcome is somewhat anticipated, given that higher grade math problems generally present greater difficulty, it is still surprising to observe that half of the models struggle even at grade 6. Among the models, DeepSeek V3 and Gemini 2.5 Flash emerge as the models achieving consistent success—defined as accuracy exceeding 80%—across all question types.

LLaMA 4 demonstrates strong performance on grade 6, but struggles with grades 7 and 8. Grock 3 succeeds in grades six and seven (accuracy exceed-ing 70%) but fails for grade eight. GPT-4 fails across all grades, having accuracy less than 70%. For certain math problems, all five models fail to produce the correct answer. Appendix A.2 presents examples where the responses from all five LLMs do not match the annotated ground truth.

---

[4]Results are obtained early June, 2025.

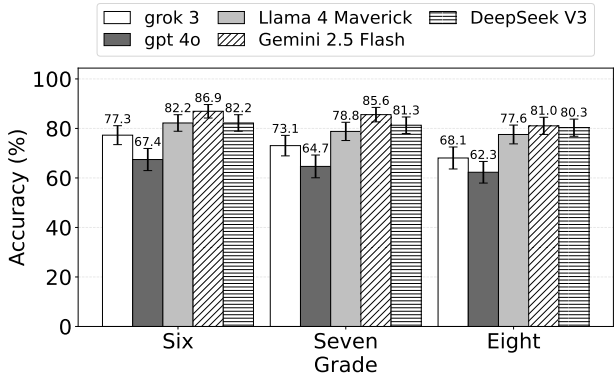

Figure 3: Average test accuracy of LLMs mathematical reasoning on the BanglaMATH dataset for grades 6, 7, and 8

Overall, the results reveal that although these math word problems are considered relatively simple for an average human adult, they continue to pose significant challenges for general-purpose open-source LLMs, particularly when presented in Bangla.

### 4.2 Arithmetic Complexity

We examine how arithmetic complexity affects the ability of LLMs to solve elementary math word problems. During annotation, we focus on arithmetic complexity of math problems, approximated by the maximum number of digits in the numbers involved. Intuitively, higher arithmetic or reasoning complexity should make problems more difficult and reduce model accuracy.

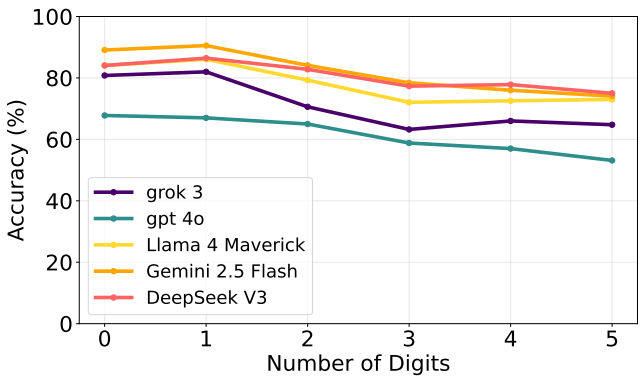

Figure 4: Average test accuracy for each LLM on the BanglaMATH dataset, based on the number of digits in the problem.

In Figure 4, we show the average test accuracy of each model on BanglaMATH, grouped by the number of digits in the problems. We observe that as the number of digits increases, the accuracy for all models drops. For problems involving zero to two digits, DeepSeek-V3, Gemini 2.5 Flash, and LLaMA 4 achieve over 80% accuracy. However, for Grok 3 and GPT-4, accuracy falls below 70% once problems require more than two-digit arithmetic. The lowest accuracies are seen with five-digit problems: among all model DeepSeek-V3 and Gemini 2.5 Flash achieving highest accuracy $\approx 75\%$ while GPT-4 accuracy lowest accuracy 58.13%. Overall, the trend clearly shows that LLMs make more mistakes as the arithmetic complexity, measured by the number of digits, increases.

### 4.3 Reasoning Complexity

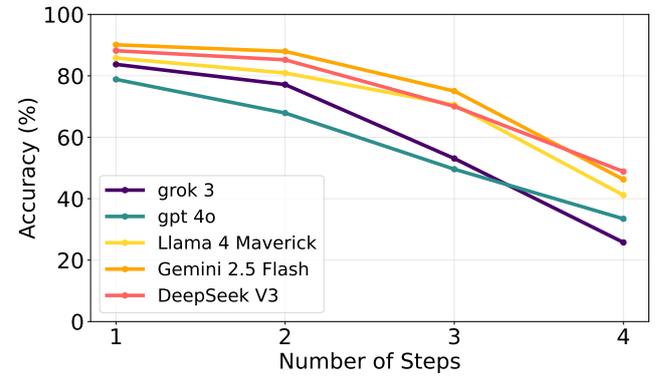

Figure 5: Average test accuracy for each LLM on the BanglaMATH dataset, based on the number of reasoning steps required to solve each problem.

To evaluate how reasoning complexity influences the ability of state-of-the-art LLMs to solve math problems, we determined the number of reasoning steps required to solve each problem during the annotation process. Naturally, problems that require a higher number of steps are considered to have greater reasoning complexity.

As shown in Figure 5, model performance drops significantly as reasoning complexity increases. When a problem requires only one or two reasoning steps, all LLMs perform consistently well, with accuracy $> 70\%$. However, for problems requiring four or more reasoning steps, the accuracy of all models falls below $< 50\%$. This pattern clearly demonstrates that LLMs are much more likely to make mistakes as reasoning complexity increases.

### 4.4 Robustness

We evaluate the robustness of the two top-performing LLMs (Gemini 2.5 Flash and DeepSeek V3) against irrelevant information. Here, irrelevant information refers to details that are related to the context of the problem but are inconsequential to its solution. This type of robustness is especially important, as real-world problems rarely appear in an idealized form where all provided information is necessary. Therefore, it is essential for LLMs to distinguish relevant from irrelevant information and use only the pertinent details to arrive at a correct solution. To test this ability, we manually construct a small distractor dataset containing 60 examples—20 from each grade level. Each example consists of an original problem along with four modified versions that include 1 to 4 pieces of irrelevant information, referred to as distractors. Each distractor includes exactly one number and integrates seamlessly into the original problem context, making it appear contextually appropriate. We test both models on this distractor dataset and observe a significant drop in performance as the number of distractors increases (see Figure 6). Both models experience an accuracy drop of approximately 20% when only two distractors are added. Notably, DeepSeek V3 suffers more degradation than Gemini 2.5 Flash when the number of distractors exceeds two.

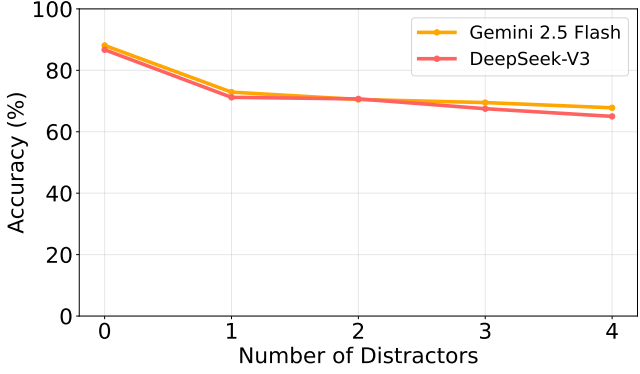

Figure 6: Test accuracy for two LLMs (Gemini 2.5 Flash and Deepseeks v3) against the number of distractors on the subset of BanglaMATH dataset.

In Appendix A.1, Table 3 shows examples of the models' responses to distractor-augmented problems and reveal that the models behave differently in the presence of irrelevant information. Both models adjust their reasoning based solely on the added irrelevant details, often leading to incorrect conclusions. Based on these results, we conclude that neither model demonstrates strong robustness to irrelevant information. This exposes a critical weakness: even advanced LLMs struggle to filter out distractions, highlighting the ongoing need for improved contextual understanding in real-world problem-solving tasks.

### 4.5 Language Bias

LLMs often exhibit decreased performance when operating in languages other than English (Dey et al., 2024). To investigate potential bias toward low-resource languages like Bangla, we conducted an ex-periment using a subset of 60 samples as distractor dataset from the BanglaMATH dataset. Each mathematical question was translated from Bangla to English, and we evaluated the performance of Gemini 2.5 Flash and DeepSeek V3 on both the original and translated versions. As shown in Figure 7, both models demonstrate a performance improvement of approximately 6.5% after translation. Notably, Gemini 2.5 Flash consistently outperforms DeepSeek V3 in both Bangla and English versions. However, the substantial accuracy gain following translation underscores a significant language bias—revealing that these LLMs reason more effectively in English than in Bangla. This disparity highlights a critical shortcoming: current LLMs trained on low-resource languages like Bangla do not yet match their English-language capabilities. Our findings emphasize the urgent need to enhance the mathematical reasoning abilities of LLMs across diverse linguistic contexts, especially for underrepresented and low-resource languages.

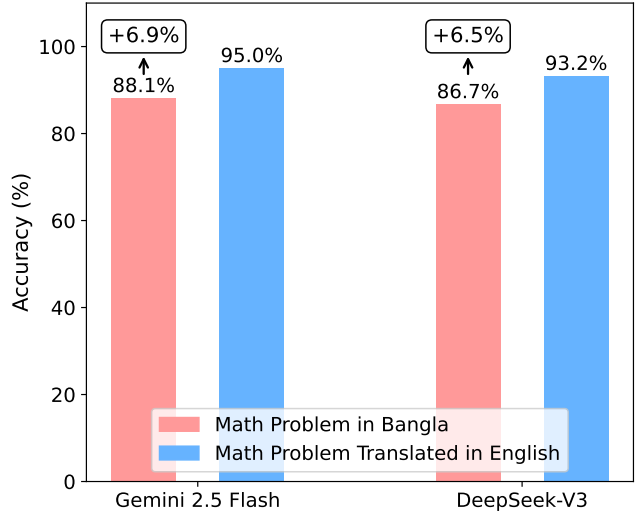

Figure 7: Test accuracy of Gemini 2.5 Flash and DeepSeek V3 on BanglaMATH subset, comparing original Bangla problems and English translations.

## 5 Related Works

Math-related datasets are available predominantly in English (Hendrycks et al., 2021; Amini et al., 2019; Cobbe et al., 2021), making them unsuitable for evaluating the reasoning abilities of LLMs in Bangla. Traditional math word problem (MWP) datasets like AddSub (Hosseini et al., 2014) and MultiArith (Roy and Roth, 2016) are integrated into broader MWP repositories. Other similar datasets include SingleEq (Koncel-Kedziorski et al., 2015), AQUA (Ling et al., 2017), and AsDiv (Miao et al., 2020). GSM8K (Clark et al., 2022) and SVAMP (Patel et al., 2021) take ad-vantage of detailed annotations and have prevailed in recent evaluations of LLMs. Similarly, MATH dataset (Hendrycks et al., 2021), which collects problems from American high school mathematics competitions and categorizes them into seven subjects: Pre-algebra, Algebra, Number Theory, Counting and Probability, Geometry, Intermediate Algebra, and Pre-calculus. However, these problems are extremely challenging—even for humans, the accuracy rate is only 40%. Given that many LLMs are still in early stages, using overly difficult problems may have limited utility in evaluating their capabilities.

Other than English, several Chinese and a Hindi (Sharma et al., 2022) math-related datasets exist. AGI-Eval (Zhong et al., 2023) and C-Eval (Huang et al., 2023) target general-purpose, multi-disciplinary evaluation for LLMs and contain subsets specifically designed to assess mathematical abilities. The math problems in these datasets range from middle school to college level and are often quite complex. Similarly, Math23K (Wang et al., 2017) and APE210K (Zhao et al., 2020), CMATH (Wei et al., 2023) and K6 (Yang et al., 2023), comprise elementary school-level math word problems. CMATH and K6, are two datasets that are relatively similar to ours and are developed concurrently. Both focus on math word problems from elementary school and organize instances by grade level.

Our work is most direclty inspired by the CMATH dataset, which contains 1.7K problems collected from online workbooks and exams, while K6 comprises 600 problems collected from an educational institution. However, neither of these two datasets has been publicly released, which prevents us from conducting an empirical comparison with them. Additionally, APE210K, contains an enormous 210K Chinese math word problems from elementary school. Its test set alone includes as many as 5,000 problems. However, the test sets do not provide annotations specific to LLM evaluation.

## 6 Conclusion

We have introduced BanglaMATH, a novel dataset designed to enable fine-grained evaluation of Large Language Models (LLMs) on elementary-level mathematical word problems in Bangla. To the best of our knowledge, BanglaMATH is the first Bangla mathematical benchmark dataset designed to help evaluate the mathematical reasoning abilities of LLMs. Our results show that BanglaMATH poses a significant challenge even for state-of-the-art LLMs, with performance declining as the grade level increases. Additionally, we observe that as the arithmetic and reasoning complexity of problems increases, the accuracy of all evaluated models decreases. Our robustness analysis reveals that top-performing models such as Gemini 2.5 Flash and DeepSeek V3 struggle when faced with irrelevant or distracting information, highlighting a vulnerability in real-world problem comprehension. Furthermore, we uncover a clear language bias—performance improves when Bangla problems are translated into English. This suggests that current LLMs are less effective at reasoning in low-resource languages like Bangla compared to English. We believe BanglaMATH can serve as a valuable benchmark for advancing non-English mathematical reasoning research, and help evaluate the cross-linguistic mathematical reasoning capabilities of LLMs.

## 7 Limitations

While we have made a strong effort to evaluate the Bangla mathematical reasoning abilities of various LLMs, due to budget constraints and restricted API access, we are unable to assess all available LLMs. We encourage future work to expand on this study by including a broader range of models, especially as more open-source and multilingual LLMs become accessible. Our focus in this work is on evaluating mathematical understanding and the results underscore that mathematical reasoning—particularly in non-English contexts and in the presence of real-world distractions—remains a significant challenge, even for state-of-the-art models. We hope that BanglaMATH serves as a valuable resource for future research aimed at improving multilingual mathematical reasoning and helps guide the development of LLMs that are robust, linguistically inclusive, and capable of solving real-world math problems effectively.

## 8 Ethics Statement

All problems in the BanglaMATH dataset are based on publicly available elementary school-level materials. No human subjects or sensitive data were involved during the annotation process; all annotations rely solely on the authors' judgment. We acknowledge that certain labels, such as the number of steps in a solution, may vary based on individual interpretation. We ensure responsible use of LLMs and adhere to ethical guidelines in data handling and model evaluation.

# A Appendix

## A.1 Model Responses Under Distractor Augmentation

Table 3: Math problem with increasing distractors and model responses from Gemini 2.5 Flash and DeepSeek V3. The "# Distractor" column shows how many irrelevant phrases were added. The first row is the original problem; subsequent rows include 1–4 distractors. The table compares how Gemini Flash 2.5 and DeepSeek V3 reasoning changes with added distractions.

| # Distractors | Math Problem | Gemini Flash 2.5 Response | DeepSeek V3 Response |
|---|---|---|---|
| 0 | খেলনা গাড়ি নিয়ে খেলার সময় মিন্টু দেখলো , সেন্টুর গাড়ির সংখ্যা তার চেয়ে বেশি এবং সেন্টু কে বললো, তুমি আমাকে দুইটা গাড়ি দাও তাহলে আমাদের গাড়ির সংখ্যা সমান হবে ।সেন্টু উল্টো জোড় করে মিন্টুর থেকে ২টি গাড়ি নিয়ে হিসাব করে বলল তার গাড়ির সংখ্যা এখন মিন্টুর দ্বিগুণ । শুরুতে কার কয়টি গাড়ি ছিল ? Mintu saw that Sentu had more toy cars than him. He said, "Give me 2 cars and we'll have the same number." Sentu instead took 2 cars from Mintu and said, "Now I have double the cars you have." What were their original counts? | **Answer:** Mintu had 10 cars, Sentu had 14. **Reasoning:** Because if Sentu gave 2 cars, both would have 12; if Sentu took 2 cars, he would have 16 and Mintu 8 — and $16 = 2 \times 8$. | **Answer:** Mintu: 10, Sentu: 14 **Reasoning:** If Sentu gives 2 cars, both have 12; if Sentu takes 2, he has 16 and Mintu 8 → $16 = 2 \times 8$ |
| 1 | "খেলনা গাড়ি নিয়ে খেলার সময় মিন্টু দেখলো , সেন্টুর গাড়ির সংখ্যা তার চেয়ে বেশি এবং সেন্টু কে বললো, তুমি আমাকে দুইটা গাড়ি দাও তাহলে আমাদের গাড়ির সংখ্যা সমান হবে ।সেন্টু উল্টো জোড় করে মিন্টুর থেকে ২টি গাড়ি নিয়ে হিসাব করে বলল তার গাড়ির সংখ্যা এখন মিন্টুর দ্বিগুণ ।তাদের মোট খেলনা গাড়ির সংখ্যা ২৫টির কম ছিল । শুরুতে কার কয়টি গাড়ি ছিল ? Mintu saw that Sentu had more toy cars than him. He said, Give me 2 cars and we'll have the same number. Sentu instead took 2 cars from Mintu and said, Now I have double the cars you have. "The total number of toy cars was less than 25 ". What were their original counts? | **Answer:** Mintu had 6 cars, Sentu had 10 **Reasoning:** Mintu initially had 6 cars and Sentu had 10 — their total is 16, which is less than 25, this is the only pair that satisfies both conditions. | **Answer:** Mintu: 10, Sentu: 14 **Reasoning:** Although the total is 24, this is the only pair that satisfies both conditions, so unchanged |

| 2 | খেলনা গাড়ি নিয়ে খেলার সময় মিন্টু দেখলো , সেন্টুর গাড়ির সংখ্যা তার চেয়ে বেশি এবং সেন্টু কে বললো, তুমি আমাকে দুইটা গাড়ি দাও তাহলে আমাদের গাড়ির সংখ্যা সমান হবে ।সেন্টু উল্টো জোড় করে মিন্টুর থেকে ২টি গাড়ি নিয়ে হিসাব করে বলল তার গাড়ির সংখ্যা এখন মিন্টুর দ্বিগুণ । তাদের মোট খেলনা গাড়ির সংখ্যা ২৫টির কম ছিল। প্রতিটি গাড়ির দাম ছিল ৩০ টাকা।শুরুতে কার কয়টি গাড়ি ছিল ? Mintu saw that Sentu had more toy cars than him. He said, Give me 2 cars and we'll have the same number. Sentu instead took 2 cars from Mintu and said, Now I have double the cars you have. " The total number of toy cars was less than 25. Each car cost 30 Taka." What were their original counts? | **Answer:** Mintu had 6 cars, Sentu had 10

**Reasoning:** The price is irrelevant; since the number conditions match, the answer remains unchanged. | **Answer:** Mintu: 10, Sentu: 14

**Reasoning:** Price or total cost does not affect the logic; the previous answer remains valid |
|---|---|---|---|
| 3 | খেলনা গাড়ি নিয়ে খেলার সময় মিন্টু দেখলো , সেন্টুর গাড়ির সংখ্যা তার চেয়ে বেশি এবং সেন্টু কে বললো, তুমি আমাকে দুইটা গাড়ি দাও তাহলে আমাদের গাড়ির সংখ্যা সমান হবে ।সেন্টু উল্টো জোড় করে মিন্টুর থেকে ২টি গাড়ি নিয়ে হিসাব করে বলল তার গাড়ির সংখ্যা এখন মিন্টুর দ্বিগুণ । তাদের মোট খেলনা গাড়ির সংখ্যা ২৫টির কম ছিল। প্রতিটি গাড়ির দাম ছিল ৩০ টাকা।তারা প্রতিদিন ৩ ঘণ্টা খেলনা গাড়ি দিয়ে খেলে। শুরুতে কার কয়টি গাড়ি ছিল ? Mintu saw that Sentu had more toy cars than him. He said, Give me 2 cars and we'll have the same number. Sentu instead took 2 cars from Mintu and said, Now I have double the cars you have. " The total number of toy cars was more than 25. Each car cost 30 Taka.They play 3 hours a day with the cars ". What were their original counts? | **Answer:** Mintu had 6 cars, Sentu had 10

**Reasoning:** playing time is irrelevant; the answer is based solely on the car count logic. Price or total cost does not affect the logic; the previous answer remains valid | **Answer:** Mintu: 10, Sentu: 14

**Reasoning:** Playing time is irrelevant; the answer is determined by the number logic alone |

| 4 | খেলনা গাড়ি নিয়ে খেলার সময় মিন্টু দেখলো , সেন্টুর গাড়ির সংখ্যা তার চেয়ে বেশি এবং সেন্টু কে বললো, তুমি আমাকে দুইটা গাড়ি দাও তাহলে আমাদের গাড়ির সংখ্যা সমান হবে ।সেন্টু উল্টো জোড় করে মিন্টুর থেকে ২টি গাড়ি নিয়ে হিসাব করে বলল তার গাড়ির সংখ্যা এখন মিন্টুর দ্বিগুণ । তাদের মোট খেলনা গাড়ির সংখ্যা ২৫টির কম ছিল। প্রতিটি গাড়ির দাম ছিল ৩০ টাকা ।তারা প্রতিদিন ৩ ঘণ্টা খেলনা গাড়ি দিয়ে খেলে।তাদের খেলনা গাড়ির সংগ্রহে ১০টি বিভিন্ন মডেল ছিল। শুরুতে কার কয়টি গাড়ি ছিল ? Mintu saw that Sentu had more toy cars than him. He said, Give me 2 cars and we'll have the same number. Sentu instead took 2 cars from Mintu and said, Now I have double the cars you have. "The total number of toy cars was more than 25. Each car cost 30 Taka.They play 3 hours a day with the cars.Their collection included 10 different models ". What were their original counts? | **Answer:**Insufficient Information<br><br>**Reasoning:** The given information in this problem makes it impossible to confirm solutions. | **Answer:**Mintu: 6, Sentu: 14<br><br>**Reasoning:** new condition (number of models) suggests a different solution |
|---|---|---|---|

## A.2 LLM Response

Table 4: Example math problems from the BanglaMATH dataset where all LLMs provide incorrect answers.

| Questions (Bangla) | Answer (Bangla) | Math Problem (English) | Answer (English) | GPT-4 | LLaMA | Gemini 2.5 Flash | Deep Seek-V3 | grok 3 |
|---|---|---|---|---|---|---|---|---|
| ১০ বছর পূর্বে মিলি ও পুষ্পার বয়সের অনুপাত ছিল ৭:১। বর্তমানে তাদের বয়সের অনুপাত ছিল ৩:১। ১৬ বছর পূর্বে মিলি ও পুষ্পার বয়সের অনুপাত কত? | ১৬ বছর পূর্বে পুষ্পার জন্ম হয়নি | 10 years ago, the ratio of Mili and Pushpa's ages was 7:1. At present, their age ratio is 3:1. What was their age ratio 16 years ago? | Pushpa was not born 16 years ago | 16:1 | 13:1 | 13:1 | 5:1 | 5:1 |
| পনির ও রবিনের আয়ের অনুপাত ৪:৩, রবিন ও তপনের অনুপাত ৫:৪। পনিরের আয় ১২০০০ টাকা হলে, তিনজনের মোট আয় কত? | ২৮২০০ টাকা | The income ratio of Ponir and Robin is 4:3, and the ratio of Robin and Topon is 5:4. If Ponir's income is 12,000 Taka, what is the total income of the three? | 28,200 Taka | 39,600 Tk. | 31,200 Tk. | 26,100 Tk. | 31,200 Tk | 22,500 Tk |
| খেলনা গাড়ি নিয়ে খেলার সময় মিন্টু দেখলো, সেন্টুর গাড়ির সংখ্যা তার চেয়ে বেশি। সেন্টুকে বললো, তুমি আমাকে দুইটা গাড়ি দাও তাহলে আমাদের গাড়ির সংখ্যা সমান হবে। সেন্টু উল্টা জোর করে মিন্টুর থেকে ২টি গাড়ি নিয়ে বলল তার গাড়ির সংখ্যা এখন মিন্টুর দ্বিগুণ। শুরুতে কার কয়টি গাড়ি ছিল? | সেন্টুর ছিল ১০টি, মিন্টুর ছিল ১৪টি খেলনা গাড়ি | While playing with toy cars, Mintu noticed that Sentu had more cars than him. Mintu said, "If you give me two cars, we'll have the same number." Sentu instead forcibly took two cars from Mintu and said, "Now I have double the cars you have." How many toy cars did each have originally? | Sentu had 10, Mintu had 14 toy cars | Mintu had 6, Shentu had 10 cars | Mintu had 14, Shentu had 18 cars | Mintu had 6, Shentu had 10 cars | Mintu had 6, Shentu had 10 cars | Mintu had 6, Shentu had 10 cars . |
| কোন একটি বিয়ের অনুষ্ঠানে রান্না করতে বাবুর্চি ও তার সহকর্মী মোট ৪০০টি পেঁয়াজ কাটেন। বাবুর্চি প্রতি মিনিটে অন্তত ৩টি পেঁয়াজ এবং তার সহকর্মী প্রতি মিনিটে অন্তত ২টি পেঁয়াজ কাটতে পারে। যদি বাবুর্চি তার সহকর্মীর চেয়ে ২৫ মিনিট আগে পেঁয়াজ কাটা বন্ধ করে, তবে কে কতটি পেঁয়াজ কেটেছিল আর কার কতক্ষণ সময় লেগেছিল? | বাবুর্চি ৭০ মিনিটে ২১০টি, সহকারী ৯৫ মিনিটে ১৯০টি | At a wedding ceremony, a chef and his assistant together cut 400 onions. The chef can cut at least 3 onions per minute, and the assistant at least 2 onions per minute. If the chef stopped cutting 25 minutes before the assistant did, how many onions did each cut and how long did they work? | Chef: 210 onions (70 min), Assistant: 190 onions (95 min) | Chef: 240 onions; Assistant: 160 onions | Chef: 240 onions; Assistant: 160 onions | Chef: 175 onions; Assistant: 240 onions | Chef: 240 onions; Assistant: 160 onions | Chef: 240 onions; Assistant: 160 onion |
| কারিম সাহেবের ৩ পুত্র এবং প্রথম ২ জন যমজ। ৩ পুত্রের বর্তমান বয়সের সমষ্টি ২৯ বছর। ৬ বছর পূর্বে তাদের বয়সের সমষ্টি ছিল ১২ বছর। পুত্রদের বর্তমান বয়স কত? | ১২, ১২, ৩ বছর | Mr. Karim has three sons, the first two are twins. The sum of their current ages is 29 years. Six years ago, the sum of their ages was 12 years. What are their current ages? | 12, 12, and 3 years old | 10,10,9 | 5,5,10 | 9,9,11 | 10,10,9 | 12,12,5 |